\documentclass[11pt]{article}

\usepackage{latexsym}
\usepackage{amsmath}
\usepackage{amssymb}

\def\punto{$\hspace*{\fill}\Box$}
\newcommand{\nop}[1]{}
\newcommand{\tuple}[1]{{\langle#1\rangle}}
\def\lBrack{\lbrack\!\lbrack}
\def\rBrack{\rbrack\!\rbrack}
\newcommand{\Bracks}[1]{\lBrack#1\rBrack}
\newcommand{\semndef}[1]{\lBrack#1\rBrack_{ndef}}
\newcommand{\algtoso}[1]{\lBrack#1\rBrack_{so}}

\def\tt{\vec{t}}
\def\pws{\mbox{subset}}

\def\rk{\mbox{repair-key}}
\def\poss{\mbox{possible}}

\newtheorem{theorem}{Theorem}[section]

\newtheorem{example}[theorem]{Example}

\newtheorem{definition}[theorem]{Definition}
\newtheorem{proposition}[theorem]{Proposition}

\newtheorem{corollary}[theorem]{Corollary}
\newtheorem{lemma}[theorem]{Lemma}
\newtheorem{remark}[theorem]{Remark}

\nop{
\addtolength{\oddsidemargin}{-1.4in}
\addtolength{\evensidemargin}{-1.4in}
\addtolength{\textheight}{-.4in}
\addtolength{\topmargin}{-1.1in}
} 

\title{A Compositional Query Algebra for Second-Order Logic and
Uncertain Databases}

\author{Christoph Koch \\ 
Department of Computer Science \\
           Cornell University, Ithaca, NY 14853, USA \\
           koch@cs.cornell.edu}
\date{}

\begin{document}

\maketitle

\begin{abstract}
World-set algebra is a variable-free query language
for uncertain databases. It constitutes the core of the query language
implemented in MayBMS, an uncertain database system.
This paper shows that world-set algebra captures exactly
second-order logic over finite structures, or equivalently,
the polynomial hierarchy.
The proofs also imply that world-set algebra is closed under
composition, a previously open problem.
\end{abstract}

\section{Introduction}

Developing suitable query languages for uncertain databases is a
substantial research challenge
that is only currently starting to get addressed.
In previous work \cite{AKO07ISQL}, we have developed a query language in the
spirit of relational algebra for processing uncertain
data -- {\em world-set algebra}\/ (WSA).
WSA consists of the operations of relational algebra plus two further
operations, one to introduce uncertainty and one to compute possible tuples
across groups of possible worlds.
WSA is implemented in the MayBMS system
\cite{AKO07ISQL,AJKO2008,KO2008,Koch2008}.

It remains to obtain an understanding of the complexity and expressive
power of world-set algebra.
The main result of this paper is a proof that world-set algebra over
uncertain databases consisting of finite sets of possible worlds
(each one a relational database) precisely
captures second-order logic (SO) over finite structures, or equivalently, the
polynomial hierarchy. This seems to be a somewhat surprising coincidence,
since the language was not designed with this result as a goal but by
abstraction from a set of use cases
from the contexts of hypothetical (``what-if'')
queries, decision support queries, and data cleaning.
Viewed differently, WSA is a natural variable-free language equivalent to
SO; it is to SO what relational algebra is
to first-order logic. To the best of the author's knowledge,
no other such language is known.

The fact that WSA exactly captures second-order logic is a strong argument
to justify it as a query language for uncertain data.
Second-order logic is a natural yardstick for languages for querying
possible worlds. Indeed, second-order quantifiers are the essence of
what-if reasoning about databases.
World-set algebra seems to be a strong candidate for a core algebra for
forming query plans and optimizing and executing them in uncertain database
management systems.

It was left open in previous work whether world-set
algebra is closed under composition, or in other words, whether
definitions are adding to the expressive power of the language.
Compositionality is a desirable and rather commonplace property of query
algebras, but in the case of WSA it seems rather unlikely
to hold. The reason for this is that the algebra contains an
uncertainty-introduction operation
that on the level of possible worlds is nondeterministic.
First materializing a view and subsequently using it multiple
times in the query is semantically quite different from composing the
query with the view and thus obtaining several copies of the view
definition that can now independently make their nondeterministic choices.
In the paper, evidence is given
that seems to suggest that definitions are essential
for the expressive power of WSA.

The paper nevertheless gives a proof that definitions do not add to the
power of the language, and WSA is indeed compositional. In fact, there
is even a (nontrivial) practical linear-time translation from SO to WSA
without definitions. This result, and the techniques for proving it,
may also be relevant in other contexts. For example, it is shown that
self-joins essentially
can always be eliminated from classical relational algebra at
the cost of introducing difference operators.

The proofs also imply that WSA is complete for the polynomial hierarchy
with respect to data complexity and PSPACE-complete with respect to
combined complexity \cite{Var82,Sto1977}.

For use as a query language for probabilistic databases, WSA has been
extended very slightly by a tuple confidence computation operation
(see e.g.\ \cite{Koch2008}).
The focus of this paper is on the nonprobabilistic
language of \cite{AKO07ISQL}.
For the efficient processing of queries of this language, the confidence
operation is naturally orthogonal to the remaining operations
\cite{AJKO2008,KO2008,Koch2008}.
The expressiveness and complexity results
obtained in the present paper constitute
lower bounds for the probabilistic version of the language. But
the non-probabilistic language is interesting and important in its own right:
Many interesting queries can be phrased in terms of the
alternatives possible in a data management scenario with uncertainty,
without reference to the relative (probability) weights of these alternatives.

The structure of this paper is as follows.
Section~\ref{sect:uncertain_so}
establishes the connection between second-order logic and uncertain
databases. Section~\ref{sect:WSA} introduces world-set algebra and gives
formal definitions of syntax and semantics.
Section~\ref{sect:WSA_SO}
proves that WSA exactly captures the expressive power of second-order
logic over finite structures. These proofs assume the availability of a
construct for making definitions (materializing views). Section~\ref{sect:weak}
discusses
the importance of being able to compose these definitions with the language,
and shows why it should seem rather surprising that definitions are not
needed for capturing second-order logic. Section~\ref{sect:composition}
finally proves that
definitions can indeed be eliminated without loss of expressive power, and
a construction for composition is given.
We obtain from these results that WSA with or without definitions
is complete for the polynomial hierarchy with respect to data complexity
and PSPACE-complete with respect to combined complexity. 
We discuss related work in Section~\ref{sect:related} and conclude in
Section~\ref{sect:conclusion}.

\section{Uncertain Databases} 
\label{sect:uncertain_so}

\def\W{W}

The schema of a relational database is a set of relation names
together with a function $sch$ that maps each relation name to a
tuple of attribute names. We use calligraphic symbols such as ${\cal A}$ for
relational databases.
The arity $|sch(R)|$ of a relation $R$ is denoted by $ar(R)$.

We will use the standard syntax of second-order logic (SO)
(see e.g.\ \cite{Lib2004}). Its semantics is defined using the
satisfaction relation $\vDash$, as usual.
Throughout this paper, we will only use second-order logic
{\em relativized}\/ to some finite set of domain elements (say, $D$),
as is common in finite model theory (cf.\ \cite{Lib2004}).
That is, first-order quantifiers $\exists x\; \phi$ are to be read as
$\exists x \; D(x) \land \phi$ and second-order quantifiers
$\exists R \; \phi$ are to be interpreted as $\exists R\; R \subseteq D^{ar(R)}
\land \phi$.

An {\em uncertain database}\/ over a given schema represents a finite set
$
\W = \{ {\cal A}_1, \dots, {\cal A}_n \}
$
of relational databases of that schema, called the
{\em possible worlds}\/.
One world among these is the true world, but we do not know which one.

A {\em representation}\/ for a finite set of possible
worlds $\W$ over schema $(R_1, \dots, R_k)$ is a pair of a relational database
schema and a formula $\omega$ over that database schema
with free second-order variables $R_1, \dots, R_k$ and
without free first-order variables such that $\omega$ is true on
exactly those structures that are in $\W$:
\[
(R_1, \dots, R_k) \vDash \omega \;\;\Leftrightarrow\;\;
(R_1, \dots, R_k) \in \W.
\]

\begin{example}[Standard Representation]
\label{ex:stdrep}
\em
Consider a representation of an uncertain database by relations 
that associate with each tuple a local condition in the form of a
conjunction of propositional literals.
A possible world is identified by a truth assignment for the propositional
variables used, and a tuple is in a possible world if the world's
truth assignment makes the tuple's clause true.

A representation database consists of a set $V$ of propositional
variables, a relation $L$ such that $L(c,p,1)$ is true iff
variable $p$ occurs positively in conjunction $c$ and 
$L(c,p,0)$ is true iff variable $p$ occurs negated in $c$,
and a representation relation $R_i'$
for each schema relation $R_i$ which
extends the schema of $R_i$ by a column to associate each
tuple with a conjunction.

Possible worlds are identified by subsets $P \subseteq V$ of variables
that are true.  A tuple $\vec{t}$ is in relation $R_i$ in possible world
$P$ if $R'_j(\vec{t}, c)$ is true
and conjunction $c$ is true for the variable assignment that makes the
variables in $P$ true and the others false.

The representation formula $\omega(R_1, \dots, R_k)$ is
\[ 
 \exists P\; P \subseteq V \land
 \bigwedge_{i=1}^k
   \forall \vec{t} \; R_i(\vec{t}) \Leftrightarrow
   \exists c\;  R'_i(\vec{t}, c) \land
        \forall p\; (L(c,p,0) \Rightarrow \neg P(p)) \land
                          (L(c,p,1) \Rightarrow      P(p)).
\] 

This is the representation system
that is essentially used in MystiQ \cite{dalvi07efficient},
Trio \cite{BDSHW2006}, and MayBMS \cite{AJKO2008}. It is
a special case of c-tables \cite{IL1984} in which local conditions are
in DNF, there is no global condition, and no variables occur in the
data tuples themselves (just in the local conditions associated with
the data tuples).
Note that it is complete in the sense that it can represent any nonempty
finite set of possible worlds. Moreover, it is succinct, i.e.,
the cardinality of the represented set of possible worlds is in
general exponential in the size of the representation database.
\punto
\end{example}

It is now easy to use 
second-order logic for expressing queries on uncertain databases encoded
by a representation. For instance, query $\phi$ is possible if
$\exists R_1 \cdots R_k \; \omega \land \phi$ and certain if
$\forall R_1 \cdots R_k \; \omega \Rightarrow \phi$.
Second-order logic allows us to use succinct representations, but also
yields very powerful hypothetical queries that can ask questions about
possible choices of {\em sets}\/ of tuples.
Such a choice of sets could be e.g. clusters of tuples in record matching
(also known as deduplication and under many other names).

\section{The Algebra}
\label{sect:WSA}

\def\AA{{\cal A}}
\def\BB{{\cal B}}
\def\cert{\mbox{certain}}
\def\vdashstar{\vdash\!\!^*\;}

\subsection{Syntax and Semantics}

World-set algebra (WSA) consists of the operations of relational algebra
(selection $\sigma$, projection $\pi$, renaming $\rho$, product $\times$,
union $\cup$, and difference $-$), two additional operations
$\rk$ and $\poss_{\vec{A}}$,
and definitions ``let $R:=Q$ in $Q'$'' where $R$ is a
new relation symbol that may be used in $Q'$.
{\em WSA without definitions}\/ is the set of WSA queries in which no
let-expressions occur.

Conceptually all operations are evaluated in each possible world individually.
The operations of relational algebra are evaluated within possible
world $\AA$ in the normal way.
Given input relation $R$, $\rk_{\vec{A}}(R)$
nondeterministically chooses a maximal repair of the functional dependency
$\vec{A} \rightarrow sch(R)$ on $R$, that is, it returns a subset $R'$ of $R$
in which $\vec{A}$ is a (super)key such that there is no superset of $R'$
which is a subset of $R$ and in which $\vec{A}$ is a (super)key.
The operation $\poss_{{\vec{A}}}(Q)$ is the only operation that can
look into alternative possible worlds. It computes,
for the current possible world given by $\AA$,
%
%
the set of possible tuples occurring in the results of $Q$ across the group of
possible worlds that agree with $\AA$ on $\pi_{\vec{A}}(Q)$.
Definitions (statements ``let $R := Q$ in $Q'$'')
extend $\AA$ by a named relation
$R$ defined by query $Q$. Since $Q$ is nondeterministic in general, the
overall set of possible worlds on which $Q'$ runs (which is relevant for
computing $\poss_{{\vec{A}}}$) may increase.

Formally, the semantics of world-set algebra is defined using a
translation $\Bracks{\cdot}^\AA_\W$
such that for a context of a set of possible worlds $\W$ and a world
$\AA \in \W$, $R$ is a possible result of world-set algebra query $Q$ iff
$R \in \Bracks{Q}^\AA_\W$:
%
%
\[
\begin{array}{rcl}
\Bracks{\{\tt\}}^\AA_\W &\;:=\;& \{\{\tt\}\}
\\ &\dots& \tt \mbox{ constant tuple}
\\[1.5ex]
\Bracks{R}^\AA_\W &:=& \{ R^\AA \}
\\[1.5ex]
\Bracks{\theta(Q)}^\AA_\W &:=&
   \{ \theta(R) \mid R \in \Bracks{Q}^\AA_\W \}
\\ &\dots& \theta \in \{ \sigma_\phi, \pi_{\vec{A}}, \rho_{A \rightarrow B} \}
\\[1.5ex]
\Bracks{Q_1 \;\theta\; Q_2}^\AA_\W &:=&
   \{ R_1 \;\theta\; R_2 \mid R_1 \in \Bracks{Q_1}^\AA_\W,
                              R_2 \in \Bracks{Q_2}^\AA_\W \}
\\ &\dots& \theta \in \{ \times, \cup, - \}
\\[1.5ex]
\Bracks{\rk_{\vec{A}}(Q)}^\AA_\W &:=&
   \{ R' \mid R' \subseteq R \in \Bracks{Q}^\AA_\W,
               \pi_{\vec{A}}(R) = \pi_{\vec{A}}(R'),
\\
&& \;\;\quad\quad \mbox{$\vec{A}$ is a key for $R_0$} \}
\\[1.5ex]
\Bracks{\poss_{{\vec{A}}}(Q)}^\AA_\W &:=&
   \big\{ \bigcup \big\{ R' \mid \BB \in W,
   R' \in \Bracks{Q}^\BB_\W, \\
&& \quad\quad\quad\quad  \pi_{\vec{A}}(R) = \pi_{\vec{A}}(R') \big\}
   \mid R \in \Bracks{Q}^\AA_\W \big\}
\\[1.5ex]
\lBrack \mbox{let } R := Q \mbox{ in } Q' \rBrack^\AA_\W &:=&
   \big\{ \Bracks{Q'}^{(\AA, R)}_{W'}
   \mid R \in \Bracks{Q}^\AA_\W \big\}
\\
\mbox{where } W'
&=& \{ (\BB, R') \mid \BB \in W,
              R' \in \Bracks{Q}^\BB_\W \}.
\end{array}
\]

Queries are run against an uncertain database $W$, and $\Bracks{Q}^{\AA}_W$
gives the result of $Q$ seen in possible world $\AA$ of $W$.
Using $\poss_\emptyset$, we can close the possible worlds semantics
and ask for possible (or, using difference, certain) tuples. For
such queries $\AA$ can be chosen arbitrarily (and the semantics function
can be considered to be of the form $\Bracks{Q}_W$).

Definitions in subexpression are unaffected by the operations higher up in the
expression tree and can be pulled to the top of the expression without
modification.
This is a direct consequence of the following fact, where we assume
that $\theta$ may be any of the WSA operations. (Thus $0 \le k \le 2$
and for $\poss_{\vec{A}}$, $k=2$.)

\begin{proposition}
\label{prop:pull_let_up}
For arbitrary WSA queries $Q$, $\theta(Q_1, \dots, Q_k)$,
if $V$ occurs only in $Q_i$,
\[
\theta(Q_1, \dots, Q_{i-1},
(\mbox{let } V := Q \mbox{ in } Q_i), Q_{i+1}, \dots, Q_k) \;=\;
\big( \mbox{let } V := Q \mbox{ in } \theta(Q_1, \dots, Q_k) \big).
\]
\end{proposition}

\noindent {\bf Proof.}
It can be shown by an easy induction that for any $Q$,
$\Bracks{Q}^{(\AA, V)}_W = \Bracks{Q}^\AA_{W'}$
where $W' = \{ \AA \mid (\AA, V) \in W \}$ if relation name $V$
does not appear in $Q$. This is immediate for all operations other
than $\poss_{\vec{A}}$.
Let $Q=\poss_{\vec{A}}(Q')$ and let the induction hypothesis hold for $Q'$,
i.e., $\Bracks{Q'}^{(\AA, V)}_W = \Bracks{Q'}^\AA_{W'}$.
Then
\begin{eqnarray*}
\Bracks{\poss_{\vec{A}}(Q')}^{(\AA, V)}_W &=&
   \Big\{ \bigcup \big\{ R' \mid (\BB, V') \in W,
   R' \in \Bracks{Q'}^{(\BB, V')}_W,
\\ && \quad\quad\quad\quad\;\;
   \pi_{\vec{A}}(R) = \pi_{\vec{A}}(R') \big\}
   \mid R \in \Bracks{Q'}^{(\AA,V)}_W \Big\}
\\
&=&
   \Big\{ \bigcup \big\{ R' \mid V' \in W',
   R' \in \Bracks{Q'}^\BB_{W'},
   \pi_{\vec{A}}(R) = \pi_{\vec{A}}(R') \big\}
   \mid R \in \Bracks{Q'}^\AA_{W'} \Big\}
\\
&=& 
\Bracks{\poss_{\vec{A}}(Q')}^\AA_{W'}.
\end{eqnarray*}

Now we apply the fact just proven to the subqueries $Q_j$ for $j \neq i$.
By definition,
\[
\Bracks{\mbox{let } V := Q \mbox{ in } \theta(Q_1, \dots, Q_k)}^\AA_{W'}
=
\{ \Bracks{\theta(Q_1, \dots, Q_k)}^{(\AA, V)}_W \mid V \in \Bracks{Q}^\AA_{W'}
\}.
\]
We distinguish between the various operations $\theta$. For relational
algebra,
\begin{eqnarray*}
\Bracks{\theta(Q_1, \dots, Q_k)}^{(\AA, V)}_W &=&
\Big\{
\theta(R_1, \dots, R_k) \mid \bigwedge_j R_j \in \Bracks{Q_j}^{(\AA, V)}_W
\Big\}
\\ &=&
\Big\{
\theta(R_1, \dots, R_k) \mid R_i \in \Bracks{Q_i}^{(\AA, V)}_W,
      \bigwedge_{j \neq i} R_j \in \Bracks{Q_j}^{\AA}_{W'}
\Big\}
\end{eqnarray*}
because $V$ only occurs in $Q_i$ and $\Bracks{Q_j}^{(\AA, V)}_W =
\Bracks{Q_j}^\AA_{W'}$ for $j \neq i$.
Thus
\begin{eqnarray*}
\Bracks{\mbox{let } V := Q \mbox{ in } \theta(Q_1, \dots, Q_k)}^\AA_{W'}
&=&
\Big\{
\theta(R_1, \dots, R_k) \mid
      \underbrace{R_i \in \Bracks{Q_i}^{(\AA, V)}_W,
                  V \in \Bracks{Q}^\AA_{W'}}_{R_i
          \in \Bracks{\mathrm{let}\; V := Q \;\mathrm{in}\; Q_i}^\AA_{W'}}, \\
&& \hspace{7mm}
      \bigwedge_{j \neq i} R_j \in \Bracks{Q_j}^{\AA}_{W'}
\Big\}
\\
&=& 
\Bracks{\theta(Q_1, \dots, Q_{i-1}, (\mbox{let } V := Q \mbox{ in } Q_i),
   Q_{i+1}, \dots, Q_k)}^\AA_{W'}
\end{eqnarray*}

The proof for the remaining operations proceeds similarly.
\punto

\medskip

In other words, they can be considered ``global''.
That is, without loss of generality
we could assume that each WSA query is of the form
\[
\mbox{let } V_1 := Q_1 \mbox{ in } ( \cdots (
\mbox{let } V_k := Q_k \mbox{ in } Q ) \cdots )
\]
where $Q$ does not contain definitions.

Observe that in the case of binary relational
algebra operations $\theta$, the set of possible worlds
$\Bracks{Q_1 \;\theta\; Q_2}^\AA_W$ is obtained by
pairing relations in the results of $\Bracks{Q_1}^\AA_W$ and
$\Bracks{Q_2}^\AA_W$. This is consistent with the intuition that
$\theta$ is applied to possible worlds $\BB$ that contain two relations
$R_1^\BB$ and $R_2^\BB$ and the result in $\BB$ is $R_1^\BB \;\theta\;
R_2^\BB$:
Proposition~\ref{prop:pull_let_up} implies that
\[
\theta(Q_1, \dots, Q_k) \;=\;
\big( \mbox{let } V_1 := Q_1, \dots,
                  V_k := Q_k \mbox{ in } \theta(V_1, \dots, V_k) \big).
\]

As a convention, we use $\{\tuple{}\}$ to represent truth and $\emptyset$
to represent falsity, over a nullary relation schema.

\begin{example}
\label{ex:color1}
\em
Given a relational database with relations
$V(V)$ and $E(\textit{From}, \textit{To})$
re\-pre\-senting a graph (directed, or undirected if $E$ is symmetric).
Then the following WSA query $Q$ returns true iff the graph is 3-colorable:
\begin{multline*}
\mbox{let}\;
R := \rk_{sch(V)}\big( V \times \rho_{C}\big(
   \{r\} \cup \{g\} \cup \{b\} \big) \big)
\;\mbox{in} \\
\mbox{possible}_\emptyset\big( \{\tuple{}\} - \pi_\emptyset(
   \sigma_{1.V=2.\textit{From} \land 2.\textit{To} = 3.V \land 1.C = 3.C}(R
      \times E \times R)) \big).
\end{multline*}
The possible relations $R$ are all the functions $V \rightarrow \{ r,g,b \}$,
and $Q$ simply asks whether there is such a function $R$ such that there
do not exist two adjacent nodes of the same color.

The corresponding SO sentence is
\[
\exists R \; \phi_{R: V \rightarrow \{r,g,b\}}
    \land \neg \exists u,v,c \; R(u,c) \land E(u,v) \land R(v,c)
\]
where $\phi_{R: V \rightarrow \{r,g,b\}}$ is a first-order sentence that
states that $R$ is a relation $\subset V \times \{r,g,b\}$ that
satisfies the functional dependency
$R: V \rightarrow \{r,g,b\}$.
\punto
\end{example}

\subsection{Derived Operations: Syntactic Sugar}

We will also consider the following operations, which are definable in the
base language:
\nop{
\[
\begin{array}{rcl@{~~}ll}
\Bracks{\pws(Q)}^\AA_\W &\vdash& R_0
   &\dots& \Bracks{Q}^\AA_\W \vdash R \supseteq R_0
\\[.9ex]
\Bracks{\mbox{choice-of}_{\vec{A}}(Q)}^\AA_\W &\vdash&
   \pi_{\vec{A} = \vec{a}}(R)
   &\dots& \Bracks{Q}^\AA_\W \vdash R,\; \vec{a} \in \pi_{\vec{A}}(R)
\\[.9ex]
\Bracks{\cert_{{\vec{A}}}(Q)}^\AA_\W &\vdash&
   \bigcap \big\{ S \mid
   \Bracks{Q}^{\BB \in \W}_\W \vdash S, \\
&& \quad\quad\;\;\;  \pi_{\vec{A}}(R) = \pi_{\vec{A}}(S) \big\}
   &\dots& \Bracks{Q}^\AA_\W \vdash R
\\[.9ex]
\Bracks{\poss(Q)}^\AA_\W &\vdash&
   \bigcup \big\{ S \mid
   \Bracks{Q}^{\BB \in \W}_\W \vdash S \big\}
\\[.9ex]
\Bracks{\cert(Q)}^\AA_\W &\vdash&
   \bigcap \big\{ S \mid
   \Bracks{Q}^{\BB \in \W}_\W \vdash S \big\}
\end{array}
\]
} 
\[
\begin{array}{rcl}
\Bracks{\pws(Q)}^\AA_\W &:=&
   \{ R' \mid R' \subseteq R \in \Bracks{Q}^\AA_\W \}
\\[1.5ex]
\Bracks{\mbox{choice-of}_{\vec{A}}(Q)}^\AA_\W &:=&
   \{ \pi_{\vec{A} = \vec{a}}(R) \mid R \in \Bracks{Q}^\AA_\W,
               \vec{a} \in \pi_{\vec{A}}(R) \}
\\[1.5ex]
\Bracks{\cert_{{\vec{A}}}(Q)}^\AA_\W &:=&
   \big\{ \bigcap \big\{ R' \mid \BB \in W,
   R' \in \Bracks{Q}^\BB_\W, \\
&& \quad\quad\quad\quad  \pi_{\vec{A}}(R) = \pi_{\vec{A}}(R') \big\}
   \mid R \in \Bracks{Q}^\AA_\W \big\}
\\[1.5ex]
\Bracks{\poss(Q)}^\AA_\W &:=&
   \big\{ \bigcup \big\{ R \mid \BB \in W,
                         R \in \Bracks{Q}^\BB_\W \big\} \big\}
\\[1.5ex]
\Bracks{\cert(Q)}^\AA_\W &:=&
   \big\{ \bigcap \big\{ R \mid \BB \in W,
                         R \in \Bracks{Q}^\BB_\W \big\} \big\}
\end{array}
\]

The operation $\pws$
nondeterministically chooses an arbitrary subset of its input relation.
The operation $\mbox{choice-of}_{\vec{A}}(R)$ nondeterministically chooses
an $\vec{a} \in \pi_{\vec{A}}(R)$ and selects those tuples $\tt$ of $R$ for
which $\tt.\vec{A} = \vec{a}$.
Conceptually, the operations $\pws$ and $\rk$ cause an exponential blowup
of the possible worlds under consideration: for instance,
on a certain database (i.e., consisting of a single possible world)
$\pws(R)$ creates the powerset of relation $R$
as the new set of possible worlds.
The operation $\cert_{{\vec{A}}}$ is the dual of $\poss_{{\vec{A}}}$
and computes those tuples common to all the worlds that agree on
$\pi_{\vec{A}}$.
The operations $\poss$ and $\cert$ compute the possible
respectively certain tuples across {\em all}\/ possible worlds.

\begin{proposition}
The operations 
$\pws$ and $\poss$ are expressible in WSA without definitions.
The operations choice-of$_{\vec{A}}$, $\cert_{{\vec{A}}}$,
and $\cert$ are definable in WSA with definitions.
\end{proposition}

\noindent {\bf Proof Sketch.}
The result is an immediate consequence of the following equi\-valences.
\[
\begin{array}{rcl@{~~}ll}
\mbox{choice-of}_{\vec{A}}(R) &=&
R \bowtie \mbox{repair-key}_{\emptyset}(\pi_{\vec{A}}(R)).
\\[.9ex]
\cert_{\vec{A}}(Q) &=& Q - \poss_{\vec{A}} \big( \poss_{\vec{A}}(Q) - Q \big)
\\[.9ex]
\pws(R) &=& \pi_{sch(R)}(\sigma_{A = 1}(\mbox{repair-key}_{sch(R)}(R \times
\rho_A(\{0,1\}))))
\\
&& \mbox{(w.l.o.g., $A \not\in sch(R)$).}
\\[.9ex]
\poss(Q) &=& \poss_{\emptyset}(Q)
\\[.9ex]
\cert(Q) &=& \cert_{\emptyset}(Q)
\end{array}
\]
The expression
$\poss_{\emptyset}(Q)$ computes the possible tuples of those worlds in
which the result of $Q$ in nonempty. But, obviously, in the remaining worlds
there are no tuples to collect. By the definition of $\cert_Q$ in terms of
$\poss_Q$, the definition of $\cert$ is correct too.
\punto

\begin{remark}
\em
The operation $\rk$ is also definable using the base operations without
$\rk$ plus $\pws$; however, such a definition seems to need
let-statements, while the definition of $\pws$ using $\rk$ does not.

In \cite{AKO07ISQL}, it was shown that the fragment obtained from WSA by
replacing $\rk$ by choice-of is a conservative extension of first-order logic.
That is, every query of that language that maps from a single possible world
to a single possible world is equivalent to a first-order query. It is
not surprising that this is not true for full WSA.
\end{remark}

\subsection{A Hypothetical Query Processing Example}

\begin{figure*}[t!]
\[
\begin{tabular}{l|cc}
Company\_Emp & C & E \\
\hline
  & $c_1$ & $e_{11}$ \\
  & $c_1$ & $e_{12}$ \\
  & $c_2$ & $e_{21}$ \\
  & $c_2$ & $e_{22}$ \\
  & $c_2$ & $e_{23}$ \\
\\
\end{tabular}
\hspace{5mm}
\begin{tabular}{l|cc}
Emp\_Skills & E & S \\
\hline
  & $e_{11}$ & $s_1$ \\
  & $e_{12}$ & $s_1$ \\
  & $e_{21}$ & $s_2$ \\
  & $e_{21}$ & $s_1$ \\
  & $e_{22}$ & $s_3$ \\
  & $e_{23}$ & $s_2$ \\
\end{tabular}
\]\[
\begin{array}{c@{~}c@{~}c@{~}c@{~}c}
&& (a) &&
\\[4mm]
\begin{tabular}{l|cc}
$U_1$ & C & E \\
\hline
  & $c_1$ & $e_{11}$ \\
\end{tabular}
&
\begin{tabular}{l|cc}
$U_2$ & C & E \\
\hline
  & $c_1$ & $e_{12}$ \\
\end{tabular}
&
\begin{tabular}{l|cc}
$U_3$ & C & E \\
\hline
  & $c_2$ & $e_{21}$ \\
\end{tabular}
&
\begin{tabular}{l|cc}
$U_4$ & C & E \\
\hline
  & $c_2$ & $e_{22}$ \\
\end{tabular}
&
\begin{tabular}{l|cc}
$U_5$ & C & E \\
\hline
  & $c_2$ & $e_{23}$ \\
\end{tabular}
\\[3mm]
&& (b) &&
\\[4mm]
\begin{tabular}{l|cc}
$V_1$ & C & E \\
\hline
  & $c_1$ & $e_{12}$ \\
\\
\end{tabular}
&
\begin{tabular}{l|cc}
$V_2$ & C & E \\
\hline
  & $c_1$ & $e_{11}$ \\
\\
\end{tabular}
&
\begin{tabular}{l|cc}
$V_3$ & C & E \\
\hline
  & $c_2$ & $e_{22}$ \\
  & $c_2$ & $e_{23}$ \\
\end{tabular}
&
\begin{tabular}{l|cc}
$V_4$ & C & E \\
\hline
  & $c_2$ & $e_{21}$ \\
  & $c_2$ & $e_{23}$ \\
\end{tabular}
&
\begin{tabular}{l|cc}
$V_5$ & C & E \\
\hline
  & $c_2$ & $e_{21}$ \\
  & $c_2$ & $e_{22}$ \\
\end{tabular}
\\[3mm]
&& (c) &&
\\[4mm]
\begin{tabular}{l|cc}
$W_1$ & C & S \\
\hline
  & $c_1$ & $s_1$ \\
\end{tabular}
&
\begin{tabular}{l|cc}
$W_2$ & C & S \\
\hline
  & $c_1$ & $s_1$ \\
\end{tabular}
&
\begin{tabular}{l|cc}
$W_3$ & C & S \\
\hline
  & $c_2$ & $s_2$ \\
\end{tabular}
&
\begin{tabular}{l|cc}
$W_4$ & C & S \\
\hline
  & $c_2$ & $s_2$ \\
\end{tabular}
&
\begin{tabular}{l|cc}
$W_5$ & C & S \\
\hline
  & $c_2$ & $s_2$ \\
\end{tabular}
\\[3mm]
&& (d) &&
\end{array}
\]
\caption{Database (a) and intermediate query results (b-d) of
         Example~\ref{ex:1}.}
\label{fig:ex1}
\end{figure*}

\begin{example}\label{ex:1}
\em
Consider the relational database of Figure~\ref{fig:ex1}(a) which
represents employees working in companies and their skills.
The query, a simplified decision support problem, will be stated in
four steps.

\begin{enumerate}
\item
{\em Suppose I choose}\/ to buy exactly one company and, as a consequence,
exactly one (key) employee leaves that company.
\[
\mbox{U} := \mbox{choice\_of}_{C, E}(\mbox{Company\_Emp})
\]
(This nondeterministically chooses a tuple from Company\_Emp.)

\item
Who are the remaining employees?
\[
\mbox{V} := \pi_{1.C, 2.E} (U
   \bowtie_{1.C = 2.C \land 1.E \neq 2.E} \mbox{Company\_Emp})
\]

\item
If I acquire that company, which skills can I obtain {\em for certain}?
\[
\mbox{W} := 
\mbox{certain}_{C}(\pi_{C, S}(V \bowtie \mbox{Emp\_Skills}))
\]
(This query computes the tuples of
$V \bowtie \mbox{Emp\_Skills}$ that are certain assuming that the company was
chosen correctly -- i.e., certain in the set of possible worlds that agree
with this world on the C column.)

\item
Now list the {\em possible}\/ acquisition targets if
the gain of the skill $s_1$ shall be guaranteed by the acquisition.
\[
\mbox{possible}(\pi_{C}(\sigma_{S = s_1}(W)))
\]
\end{enumerate}

Figure~\ref{fig:ex1}(b-d) shows the development of the uncertain database
through steps 1 to 3. The first step creates five possible worlds
corresponding to the five possible choices of company and renegade employee
from relation Company\_Emp. Steps two to four further process the query,
and the overall result, which is the same in all five possible worlds, is
\[
\begin{tabular}{l|c}
Result & C \\
\hline
  & $c_1$ \\
\end{tabular}
\]
\punto
\end{example}

\section{WSA with Definitions Captures SO Logic}
\label{sect:WSA_SO}

In this section, it is shown that WSA with definitions has exactly the
same expressive power as second-order logic over finite structures.

\begin{theorem}
\label{theo:SO2WSAwdef}
For every SO query, there is an equivalent WSA query with definitions.
\end{theorem}

\noindent {\bf Proof.}
%
%
We may assume without loss of generality that the SO query is a first-order
query prefixed by a sequence of second-order quantifiers.
The proposition follows from induction.

Induction start: FO queries can be translated to relational
algebra by a well-known translation known in the database context as
one direction of Codd's Theorem (cf.\ \cite{AHV95}).

Induction step (second-order existential quantification,
$\exists R_{k+1} (\subseteq D^l)\; \phi$):
Let $\phi$ be an SO formula with free second-order variables
$R_1, \dots, R_{k+1}$ and free first-order variables $\vec{x}$ where
$R_{k+1}$ has arity $l$.
Let $Q_\phi$ be an equivalent WSA expression.
Without loss of generality,
we may assume that the relations $R_1, \dots, R_k, Q_\phi$
have disjoint schemas. Let
\[ 
Q := (\mbox{let } R_{k+1} := \pws(D^l) \mbox{ in } 
\pi_{sch(Q)}(
\poss_{sch(R_1) \dots sch(R_k)}(1_{R_1} \times \dots \times 1_{R_k} \times Q_\phi))).
\] 
where
$
1_{R_i} = R_i \times \{1\} \cup (D^{ar(R_i)} - R_i) \times \{0\}. 
$
(Note that the relations $1_{R_i}$ will play a prominent role in later
parts of this paper.)
We prove that
\[
(R_1, \dots, R_k, \vec{x}) \vDash \exists R_{k+1} (\subseteq D^l) \; \phi
\;\;\Leftrightarrow\;\;  \vec{x} \in R_Q
\]
where $\{ R_Q \} = \Bracks{Q}^{(R_1, \dots, R_k)}_W$.
By definition of $\Bracks{\cdot}$,
\[
\Bracks{Q}^{(R_1, \dots, R_k)}_W =
\big\{
\pi_{sch(Q_\phi)}(\Bracks{Q'}^{(R_1, \dots, R_k+1)}_{W'}) \mid
   R_{k+1} \subseteq D^l
\big\}
\]
where $W' = \{ (R_1, \dots, R_{k+1}) \mid (R_1, \dots, R_k) \in W,
   R_{k+1} \subseteq D^l \}$ and $Q'$ is a shortcut for
$\poss_{sch(R_1) \dots sch(R_k)}(1_{R_1} \times \dots \times 1_{R_k} \times
Q_\phi)$.

We may assume a nonempty domain $D$, so
the result of $1_{R_1} \times \dots \times 1_{R_k}$ is never empty,
the mapping $(R_1, \dots, R_k) \mapsto 
1_{R_1} \times \dots \times 1_{R_k}$ is injective, and
$Q$ will therefore group the possible outcomes of $Q_\phi$ for the
various choices of $R_{k+1}$ by $R_1, \dots, R_k$.

Formally, by definition of $\Bracks{\cdot}$,
\begin{eqnarray*}
\Bracks{Q'}^{(R_1, \dots, R_{k+1})}_{W'} &=&
\Big\{ \bigcup \big\{
1_{R_1} \times \dots \times 1_{R_k} \times
\Bracks{Q_\phi}^{(R_1, \dots, R_k, R_{k+1}')}_{W'} \; \mid \\
&& \quad\quad
(R_1, \dots, R_k, R_{k+1}') \in W' \big\} \mid 
(R_1, \dots, R_{k+1}) \in W' \Big\}
\\ &=&
\Big\{ 1_{R_1} \times \dots \times 1_{R_k} \times
\bigcup \big\{
\Bracks{Q_\phi}^{(R_1, \dots, R_k, R_{k+1}')}_{W'} \mid
R_{k+1}' \subseteq D^l \big\} \Big\}.
\end{eqnarray*}
Thus, in a given world $(R_1, \dots, R_k)$, $Q$ produces exactly one world
as the result,
\[
\Bracks{Q}^{(R_1, \dots, R_k)}_W =
\Big\{ \bigcup \big\{ \Bracks{Q_\phi}^{(R_1, \dots, R_k, R_{k+1}')}_{W'} \mid
R_{k+1}' \subseteq D^l \big\} \Big\} = \{ R_Q \}
\]
and this captures exactly second-order existential quantification.

The WSA expression for universal second-order quantifiers
$\forall R_{k+1} (\subseteq D^l) \; \phi$ is similar.
Alternatively, $\forall R_{k+1}\; \phi$ can also be taken as
$\neg \exists R_{k+1}\; \neg \phi$, where complementation with
respect to $D$ is straightforward using the difference operation.
\punto

\begin{example}
\label{ex:QBF1}
\em
$\Sigma_2$-QBF
is the following $\Sigma^P_2$-com\-plete decision problem.
Given two disjoint sets of propositional variables $V_1$ and $V_2$ and
a DNF formula $\phi$ over the variables of $V_1$ and $V_2$,
does there exist a truth assignment for the variables $V_1$
such that $\phi$ is true for all truth assignments for the variables $V_2$?

Instances of this problem shall be represented
by sets $V_1$ and $V_2$, a set $C$ of ids of clauses in $\phi$, and
a ternary relation $L(C,P,S)$ such that
$\tuple{c,p,1} \in L$ (resp., $\tuple{c,p,0} \in L$)
iff propositional variable $p$ occurs
positively (resp., negatively) in clause $c$ of $\phi$, i.e.,
\[
\phi = \bigvee_{c \in C} \bigwedge_{\tuple{c,p,1} \in L} p \land
                         \bigwedge_{\tuple{c,p,0} \in L} \neg p.
\]
The QBF is true iff second-order sentence
\[
\exists P_1 \; (P_1 \subseteq V_1) \land \forall P_2 \;
    (P_2 \subseteq V_2) \Rightarrow \psi
\]
is true, where $\psi$ is the first-order sentence
\[
\exists c \; \neg \exists p \;
\big( L(c,p,0) \land      (P_1(p) \lor P_2(p)) \big) \lor
\big( L(c,p,1) \land \neg (P_1(p) \lor P_2(p)) \big).
\]
which asserts the truth of $\phi$:
that there is a clause $c$ in $\phi$ of which no literal is
inconsistent with the truth assignment
$p \mapsto (p \in P_1 \cup P_2)$.
By Theorem~\ref{theo:SO2WSAwdef},
this can be expressed as the Boolean WSA query
\begin{multline*}
\mbox{let}\; P_1 := \pws(V_1) \;\mbox{in}\;
\mbox{possible} \big( \{\tuple{}\} \\
- \;
\mbox{let}\; P_2 := \pws(V_2) \;\mbox{in}\;
\mbox{possible}_{sch(P_1)}(1_{P_1} \times (\{\tuple{}\} - Q) ) \big)
\end{multline*}
where
\[
Q = \pi_\emptyset \big( C - \pi_C \big(
(\sigma_{S=0}(L) \bowtie (P_1 \cup P_2)) \; \cup
   (\sigma_{S=1}(L) \bowtie ((V_1 \cup V_2) - (P_1 \cup P_2))) \big) \big)
\]
is relational algebra for $\psi$.
\punto
\end{example}

For the converse result, we must first make precise how second-order
logic will be compared to WSA, since second-order logic queries are usually not
``run'' on uncertain databases.
%
%
We will consider WSA queries that are evaluated
against a (single-world) relational database $\AA$
representing an uncertain database (e.g., using the standard representation
of Example~\ref{ex:stdrep}).
We already know that
arbitrary uncertain databases (that is, nonempty finite sets of possible
worlds) can be so represented, and this
assumption means no loss of generality.
The query constructs the uncertain database
from the representation and is always evaluated
as $\Bracks{Q}^{\AA}_{\{\AA\}}$, precisely as sketched at the end of
Section~\ref{sect:uncertain_so}.

\begin{theorem}
\label{theo:wsa2so}
For every WSA query, there is an equivalent second-order logic query.
\end{theorem}

\noindent {\bf Proof Sketch.}
The proof revolves around the definition of a function $\algtoso{\cdot}$
that maps each WSA expression $Q$ to an SO formula $\algtoso{Q}$
with free second-order variables $\vec{R}$ and $R_Q$
and without free first-order
variables such that $\algtoso{Q}$ and $Q$ are equivalent in the sense that
$\algtoso{Q}$ is true on structure
$(\AA, \vec{R}, R_Q)$ iff $R_Q$ is among the possible
results of $Q$ starting from possible world $(\AA, \vec{R})$.
We can state this notion of correctness,
which is the hypothesis of the following induction along the structure
of the WSA expression, formally as
\[
(\AA, \vec{R}, R_Q) \vDash \algtoso{Q} \;\;\Leftrightarrow\;\;
R_Q \in \Bracks{Q}^{(\AA, \vec{R})}_W
\]
for
\[
W = \Big\{ (\AA, \vec{R}) \mid
           (\AA, \vec{R}) \vDash
   \bigwedge_{V \;\mathrm{in}\; \vec{R}} \psi_V \Big\}.
\]
Here the free second-order variables $\vec{R}$ are also the names of
the views defined (using let-expressions)
along the path from the root of the parse tree of the query
to the subexpression $Q$.
A formula $\psi_V$ is identified by the name
of the view relation $V$, assuming without loss of generality
that each view name is
introduced only once by a let expression across the entire query.
The formulae $\psi_V$ will be defined below.

For the operations $\theta$ of relational algebra,
\begin{multline*}
\algtoso{\theta(Q_1, \dots, Q_{ar(\theta)})}(\vec{R}, R_Q) :=
   \exists R_{Q_1} \cdots R_{Q_{ar(\theta)}} \;
      \Big( \bigwedge_{i=1}^{ar(\theta)}
         \algtoso{Q_i}(\vec{R}, R_{Q_i}) \Big) \\
\land\;
      \forall \vec{x}\; R_Q(\vec{x}) \Leftrightarrow
      \phi_{\theta(Q_1, \dots, Q_{ar(\theta)})}(\vec{x})
\end{multline*}
where $0 \le ar(\theta) \le 2$ and
$
\phi_{S}(\vec{x}) := S(\vec{x})
$, where $S$ is either a relation from $\AA$ or a second-order variable
from $\vec{R}$,
$
\phi_{\{\vec{t}\}}(\vec{x}) := \vec{x} = \vec{t}
$, $
\phi_{Q_1 \cup Q_2}(\vec{x}) := R_{Q_1}(\vec{x}) \lor R_{Q_2}(\vec{x})
$, $
\phi_{Q_1 - Q_2}(\vec{x}) := R_{Q_1}(\vec{x}) \land \neg R_{Q_2}(\vec{x})
$, $
\phi_{Q_1 \times Q_2}(\vec{x}, \vec{y})
   := R_{Q_1}(\vec{x}) \land R_{Q_2}(\vec{y})
$, $
\phi_{\sigma_\gamma(Q)}(\vec{x}) := R_{Q}(\vec{x}) \land \gamma
$, $
\phi_{\pi_{\vec{x}}(Q)}(\vec{x}) := \exists \vec{y} \; R_{Q}(\vec{x}, \vec{y})
$, and $
\phi_{\rho_{\vec{x} \rightarrow \vec{y}}(Q)}(\vec{y})
   := \exists \vec{x}\; R_{Q}(\vec{x}) \land \vec{x} = \vec{y}$.
It is easy to verify that for any tuple $\vec{x}$
and relational algebra operation $\theta$,
$
(\AA, R_{Q_1}, \dots, R_{Q_{ar(\theta)}}) \vDash
   \phi_{\theta(Q_1, \dots, Q_{ar(\theta)})}(\vec{x})
$
if and only if $\vec{x}$ is a result tuple of relational algebra query
$\theta(R_{Q_1}, \dots, R_{Q_{ar(\theta)}})$.
Assume that the induction hypothesis holds for the subqueries
$Q_1, \dots, Q_{ar(\theta)}$, i.e.,
$(\AA, \vec{R}, R_{Q_i}) \vDash \algtoso{Q_i}$ if and only if
$R_{Q_i} \in \Bracks{Q_i}^{(\AA, \vec{R})}_W$ for $1 \le i \le ar(\theta)$.
The formula $\algtoso{\theta(Q_1, \dots, Q_{ar(\theta)})}$
just states that $R_Q$ is a
relation consisting of exactly those tuples $\vec{x}$ that satisfy
$\phi_{\theta(Q_1, \dots, Q_{ar(\theta)})}(\vec{x})$ for a choice of possible
results $R_{Q_i} \in \Bracks{Q_i}^{(\AA, \vec{R})}_W$ of the subqueries
$Q_i$, for $1 \le i \le ar(\theta)$.
But this is exactly the definition of
$
\Bracks{\theta(Q_1, \dots, Q_{ar(\theta)})}^{(\AA, \vec{R})}_W.
$

This in particular
covers the nullary operations of relational algebra ($\{\vec{t}\}$ and
$R$), which form the induction start.

The remaining operations are those special to WSA (with definitions):
\begin{eqnarray*}
\algtoso{\pws(Q_1)}(\vec{R}, R_Q)
   &:=& \exists R_{Q_1} \; \algtoso{Q_1}(\vec{R}, R_{Q_1}) \land
           R_Q \subseteq R_{Q_1}
\\[1ex]
\algtoso{\rk_{\vec{A}}(Q_1)}(\vec{R}, R_Q)
   &:=& \exists R_{Q_1} \; \algtoso{Q_1}(\vec{R}, R_{Q_1}) \land
         R_Q \subseteq R_{Q_1} \\
   &\land& \;\;\mbox{$\vec{A}$ is a key for $R_Q$}
\\
   &\land&
        \neg \exists R_Q' \; R_Q \subset R_Q' \subseteq R_{Q_1} \land
        \mbox{$\vec{A}$ is a key for $R_Q'$}
\\[1ex]
\algtoso{\mbox{let } V := Q_1 \mbox{ in } Q_2}(\vec{R}, R_Q)
   &:=& \exists V \; \psi_V \land
        \algtoso{Q_2}(\vec{R}, V, R_Q)
\\ && \mbox{and define } \psi_V := \algtoso{Q_1}(\vec{R}, V) 
\\[1ex]
\algtoso{\poss_{\vec{A}}(Q_1)}(\vec{R}, R_Q)
   &:=& \exists R_{Q_1} \; \algtoso{Q_1}(\vec{R}, R_{Q_1})
  \land \; \forall \vec{x}\; R_Q(\vec{x}) \Leftrightarrow \\
&& \exists \vec{R}\; \Big( \Big( \bigwedge_{V \; \mathrm{in} \; \vec{R}} \psi_V \Big) \\
&& \quad\;\; \land \; \exists R_{Q_1}' \; \algtoso{Q_1}(\vec{R}, R_{Q_1}')
\\
&& \quad\;\; \land \; \pi_A(R_{Q_1}) = \pi_A(R_{Q_1}') \land R_{Q_1}'(\vec{x})
\Big)
\end{eqnarray*}
where ``$\vec{A}$ is a key for $R$'' and
$\pi_{\vec{A}}(\cdot) = \pi_{\vec{A}}(\cdot)$ are easily expressible in FO.

It is straightforward to verify the correctness of $\algtoso{\cdot}$ for
$\pws$ and $\rk$: The definitions of $\algtoso{\cdot}$ and
$\Bracks{\cdot}$ essentially coincide.

Similarly, the correctness of the
definition of $\algtoso{\cdot}$ for let is easy to verify. Here we also
define the formulae $\psi_V$.

Finally, $\algtoso{\poss_{\vec{A}}(Q_1)}$ makes reference to world-set $W$
and for that purpose uses the formulae $\psi_V$: Indeed, the worlds in $W$
are exactly those structures that satisfy all the $\psi_V$ for relations
$V$ defined by let expressions on the path from the root of the query to the
current subexpression $\poss_{\vec{A}}(Q_1)$.
The definition 
$\algtoso{\poss_{\vec{A}}(Q_1)}$ is again very close to the definition of
$\Bracks{\poss_{\vec{A}}(Q_1)}$, and its correctness is straightforward to
verify.

Note that by eliminating the definitions $\psi_V$ we in general obtain
an exponential-size formula.
\punto

\section{Intermezzo: Why we are not done}
\label{sect:weak}

The proof that WSA with definitions can express any SO query
may seem to settle the expressiveness question for our language.
However, understanding WSA without definitions is also important, for two
reasons. First, it is a commonplace and desirable property of query algebras
that they be compositional, i.e., that the power to define views is not
needed for the expressive power, and all views can be eliminated by composing
the query. Second, if this property does not hold, it means that in general we
have to precompute and materialize views. And indeed, superficially we
would expect that WSA is not compositional in that respect: it supports
nondeterministic operations ($\rk$ and/or $\pws$). If a view definition
$V$ contains such a nondeterministic operation and a query uses
$V$ at least twice, replacing each occurrence with the definition will
not be equivalent because the two copies of the definition of $V$ will
produce different relations in some worlds.
For example, $(\mbox{let } V := \pws(U) \mbox{ in } V \bowtie V)$ is not
at all equivalent to $\pws(U) \bowtie \pws(U)$.

The question remains whether for each WSA query there is an equivalent
query in WSA without definitions via a less direct rewriting.
The answer to this question is less obvious. Our language definition
has assumed $\rk$ to be the base operation and $\pws$ definable using WSA
with $\rk$. Indeed, in WSA with definitions, either one can be defined
using the other. However, it can be shown that $\rk$ cannot be expressed
using $\pws$ without using definitions even though $\pws$ can guess
subsets and appears comparable in expressiveness to $\rk$.

\medskip

Consider possible worlds databases in which each relation is independent from
the other relations, i.e., the world set is of the form
\[
\{ (R_1, \dots, R_k) \mid R_1 \in W_1, \dots, R_k \in W_k \}.
\]
WSA without definitions on such {\em relation-independent databases}\/
gives rise to a much simpler and more intuitive semantics
definition than the one of Section~\ref{sect:WSA},
via the following function $\semndef{\cdot}$.
\begin{eqnarray*}
\semndef{\theta}(W_1, \dots, W_{ar(\theta)}) &\;:=\;&
   \{ \theta(R_1, \dots, R_{ar(\theta)}) \mid
       R_1 \in W_1, \dots, R_{ar(\theta)} \in W_{ar(\theta)} \}
\\
&& \dots \mbox{ where $\theta$ is an operation of relational algebra}
\\[1ex]
\semndef{\rk_{\vec{A}}}(W) &:=& \{ R \mid R \subseteq R' \in W,
   \pi_A(R) = \pi_A(R'), \mbox{$\vec{A}$ is a key for $R$} \}
\\[1ex]
\semndef{\pws}(W) &:=& \{ R \mid R \subseteq R' \in W \}
\\[1ex]
\semndef{\poss_{\vec{A}}}(W) &:=&
   \Big\{ \bigcup \{ R' \in W \mid \pi_{\vec{A}}(R) = \pi_{\vec{A}}(R') \}
      \mid R \in W \Big\}
\end{eqnarray*}

The correctness of this alternative semantics definition, stated next,
is easy to verify.

\begin{proposition}
For relation-independent databases and WSA without definitions,
$\semndef{\cdot}$ is equivalent to
$\Bracks{\cdot}$ in the sense that for any operation $\theta$,
\[
\{ \Bracks{\theta(Q_1, \dots, Q_{ar(\theta)})}^\AA_W \mid \AA \in W \} =
\semndef{\theta}(W_1, \dots, W_{ar(\theta)})
\]
where
$
W_i = \bigcup \{ \Bracks{Q_i}^\AA_W \mid \AA \in W \}
$
for all $1 \le i \le ar(\theta)$.
\end{proposition}

The following result asserts that adding $\pws$ to relational algebra
yields little expressive power.
By the existence of supremum of a set of worlds $W$, we assert the
existence of an element $(\bigcup W) \in W$, denoted 
$\mbox{sup}(W)$. An infimum is a set
$\mbox{inf}(W) := (\bigcap W) \in W$.

\begin{theorem}
\label{theo:pws_ncapture}
Any world-set computable using relational algebra extended by the operation
$\pws$ has a supremum and an infimum.
\end{theorem}

\noindent {\bf Proof.}
The nullary relational algebra expressions ($\{\vec{t}\}$ and $R$)
yield just a singleton world-set, and the
single world is both the supremum and the infimum.
Given a world-set $W$, $\mbox{sup}(\semndef{\pws}(W)) := \mbox{sup}(W)$ and
$\mbox{inf}(\semndef{\pws}(W)) := \emptyset$.
For a positive relational algebra expression $\theta$,
$\mbox{sup}(\semndef{\theta}(W_1, \dots, W_k)) :=
\theta(\mbox{sup}(W_1), \dots, \mbox{sup}(W_k))$ and
$\mbox{inf}(\semndef{\theta}(W_1, \dots, W_k)) :=
\theta(\mbox{inf}(W_1), \dots, \mbox{inf}(W_k))$.
For relational difference, it can be verified that
$\mbox{sup}(\semndef{-}(W_1, W_2)) := \mbox{sup}(W_1) - \mbox{inf}(W_2)$ and
$\mbox{inf}(\semndef{-}(W_1, W_2)) := \mbox{inf}(W_1) - \mbox{sup}(W_2)$.
It is easy to verify the correctness of these definitions, and together they
yield the theorem.
\punto


\medskip

Thus, not even $\rk_\emptyset(\{0,1\}) = \big\{ \{0\}, \{1\} \big\}$
can be defined.

\begin{corollary}
The set of worlds $\big\{ \{0\}, \{1\} \big\}$ is not definable
in relational algebra extended by $\pws$.
\end{corollary}

\nop{
\noindent {\bf Proof Sketch.}
Assume $\rk_{\emptyset}(\{0,1\})$
is expressible by query $Q$ on an empty schema and a single possible world.
Then $Q$ contains a subexpression
$\pws(Q_0)$ for some $\pws$-free subquery $Q_0$. It follows that
$Q_0$ is constant and a single possible world.
Let $U$ be the relation returned by $Q_0$.
Take any two $R^1, R^2 \subseteq U$ with
$R^1 \neq R^2$. There are two possible worlds in which this subexpression
$\pws(Q_0)$ has chosen $R^1$ and $R^2$, respectively, and all other $\pws$
operations of $Q$
have chosen their input relation (the maximal set they could
choose), i.e., consider all relations and $\pws$-choices other than
$R^1$ and $R^2$
fixed. Let $Q^1$ and $Q^2$ denote the results of $Q$ in these two possible
worlds. Then by Proposition~\ref{prop:mono}, either $Q^1 \subseteq Q^2$
or $Q^1 \supseteq Q^2$. Since no two possible results of a $\rk$ operation
strictly contain each other, we must conclude that $Q^1 = Q^2$. But then,
since $R$ and $R'$ were chosen arbitrarily, the output of $Q$ must not
depend on the choice made by $\pws(Q_0)$ and we can simply replace it by
$Q_0$. By repeating this argument, we see that $Q$ is equivalent to a
query without $\pws$ statements. Clearly, on a single-world input, this
returns a single possible world, which is in contradiction with our
assumption that $Q$ expresses a $\rk$ expression.
\punto
} 

In contrast, $\rk_{sch(U)}(U \times \{0,1\})$
can be defined as follows in the language fragment of relational algebra
plus $\pws$ if definitions are available:
\[
\mbox{let } R := \pws(U) \mbox{ in } (R \times \{1\} \cup (U-R) \times \{0\}).
\]

Thus, removing definitions seems to cause a substantial reduction
of expressive power. In the remainder of this paper, we study whether
$\poss_{\vec{A}}$ and $\rk$ can offset this.

Before we move on, another simple result shall be stated that gives an
intuition for the apparent weakness of WSA without definitions.
If a view is defined by a query that involves one of the
nondeterministic operations ($\poss_{\vec{A}}$ or $\rk$), then this
view can only be used at one place in the query if the query is to be
composed with the view. However, subsequent relational algebra operations
will be monotonic with respect to that view.

\begin{proposition}
\label{prop:mono}
Let $Q$ be a nonmonotonic relational algebra query that is built using a
relation $R$ and constant relations.  Then $R$ occurs at least twice in $Q$.
\end{proposition}

\noindent {\bf Proof.}
Assume a relational algebra query tree exists that expresses $Q$ and in which
$R$ only occurs as a single leaf. Then the path from that leaf towards
the root operation consists of unary operations and operations
$Q_1 \;\theta\; Q_2$ where $Q_1$ contains $R$ and $Q_2$ has only constant
relations as leaves: $Q_2$ is constant. So $Q_1 \;\theta\; Q_2$
can be thought of as a unary operation.
But all unary operations $\theta$ are monotonic, i.e., if
$X \subseteq Y$, then $\theta(X) \supseteq \theta(Y)$
for the family of operations $(C-X)_{C \;\mathrm{const.}, sch(C)=sch(X)}$
and $\theta(X) \subseteq \theta(Y)$ for all other operations.
It follows that $Q$, a sequence of such operations, is also monotonic.
\punto

\section{WSA without Definitions Expresses all of SO Logic}
\label{sect:composition}

As the main technical result of the paper, we now show that WSA without
definitions (but using $\rk$ as in our language definition), captures
all of SO. It follows that definitions, despite our nondeterministic operations,
do not add power to the language. This is surprising given
Theorem~\ref{theo:pws_ncapture}.

\subsection{Indicator Relations}
\label{subsect:ind}

\def\compl{\mathrm{compl}}
\def\nonempty{\mathrm{nonempty}}
\def\makeind{\mathrm{ind}}

Let $U$ be a {\em nonempty}\/
relation (the {\em universe}\/) and let $R \subseteq U$.
Then the indicator function $1_R: U \rightarrow \{0,1\}$ is defined as
\[
1_R: x \mapsto \left\{
\begin{array}{lll}
1 & \dots & x \in R \\
0 & \dots & x \not\in R \\
\end{array}
\right.
\]
The corresponding indicator relation is just the relation
$
\{ \tuple{x, 1_R(x)} \mid x \in U \}
$
which, obviously, has functional dependency $U \rightarrow \{0,1\}$.
Subsequently, we will always use indicator relations rather than indicator
functions and will denote them by $1_R$ as well.
%
%
By our assumption that $U \neq \emptyset$, indicator relations
are always nonempty.

Given relations $R$ and $U$ with $R \subseteq U \neq \emptyset$,
the indicator relation $1_R$ w.r.t.\ universe $U$
can be constructed in relational algebra as
\[
\makeind(R,U) := (R \times \{1\}) \cup ((U-R) \times \{0\}).
\]

The expression
$\rk_{sch(U)}(U \times \{0,1\})$ is equivalent to
\[
\mbox{let $R := \pws(U)$ in} \; \makeind(R,U)
\]
and yields an indicator relation in each possible world.

Indicator relations have the nice property that their complement
can be computed using a conjunctive query (with an inequality),
\[
1_{U-R} = (U \times \{0,1\}) - 1_R  :=
\pi_{1,2}(\sigma_{1=3 \land 2 \neq 4}(U \times \{0,1\} \times 1_R)).
\]

Let $\overline{R}$ denote the complement of relation $R$ and
let $U_i = R_i \cup \overline{R_i}$, called the {\em universe}\/ of $R_i$.
Note that
\[
\overline{R_1 \times \dots \times R_k} =
   \bigcup_{i=1}^k U_1 \times \dots \times U_{i-1} \times \overline{R_i}
      \times U_{i+1} \times \dots \times U_k.
\]

The complement of a product
$\vec{1} := 1_{R_1} \times \dots \times 1_{R_k}$
can be obtained as
\begin{eqnarray*}
\compl_{U_1,\dots,U_k}(\vec{1}) &=&
   (U_1 \times \{0,1\} \times \dots \times U_k \times \{0,1\})
   - \vec{1}
\\
&=& \pi_{A_1,B_1,\dots,A_k,B_k} (
    \sigma_{\bigvee_i (A_i=A'_i \land B_i \neq B'_i)} (
       \rho_{A'_1 B'_1 \dots A'_k B'_k}(\vec{1}) \;\times
\\
&& \hspace{5mm}
       \rho_{A_1 B_1 \dots A_k B_k} ( 
           U_1 \times \{0,1\} \times \dots \times U_k \times \{0,1\} )
) ).
\end{eqnarray*}
if, for each $1 \le i \le k$, $U_i$ is the universe of $R_i$.
Moreover,

\begin{lemma}
\label{lem:kprod}
The $k$-times product of $1_R$, denoted by
$(1_R)^k_U := \overbrace{1_R \times \dots \times 1_R}^{\mbox{$k$ times}}$,
can be expressed as a relational
algebra expression in which $1_R$ only occurs once.
\end{lemma}

\noindent {\bf Proof.}
Let $U$ be the universe of $R$.
\begin{eqnarray*}
(1_R)^k_U
        &=&  \rho_{A_1 B_1 \dots A_k B_k}((U \times \{0,1\})^k)
             - \compl_{U^k}(1_R^k) \\
        &=&  \rho_{A_1 B_1 \dots A_k B_k}((U \times \{0,1\})^k) \\
        &-& 
    \pi_{A_1,B_1,\dots,A_k,B_k}(
       \sigma_{\bigvee_{1 \le i \le k}(A_1=A' \land B_i \neq B')}(
\\
&& \hspace{3mm}
          \rho_{A_1 B_1 \dots A_k B_k}((U \times \{0,1\})^k)
          \times \rho_{A' B'}(1_R))).
\end{eqnarray*}
\punto

\medskip

As a convention, let $S^0 = \{\tuple{}\}$ for nonempty relations $S$.
In particular, $(1_R)^0_U = \{\tuple{}\}$.

\subsection{The Quantifier-Free Case}

By quantifier-free formulae we will denote formulae of predicate logic that
have neither first- nor second-order quantifiers.

\begin{lemma}
\label{lem:normal_form}
Let $\phi$ be a quantifier-free formula with relations $\vec{R}$.
Then $\phi$ can be translated in linear time into a
formula $\exists \vec{x} \; \alpha \land \beta$,
where $\alpha$ is a Boolean combination of equalities and
$\beta$ is a conjunction of relational literals,
which is equivalent to $\phi$
on structures in which each relation of $\vec{R}$
and its complement are nonempty.
\end{lemma}

\noindent {\bf Proof Sketch}.
Let $R_1, \dots, R_s$ the set of distinct predicates (relation names) occurring
in $\phi$.
First push negations in $\phi$ down to the atomic formulae using De Morgan's
laws and the elimination of double negation and replace relational
atomic formulae $\neg R_j(\vec{t})$,
where $\vec{t}$ is a tuple of variables and constants,
by $\overline{R_j}(\vec{t})$.

Now apply the following translation inductively bottom-up.
The translation is the identity on inequality literals.
Rewrite atomic formulae $R_j(\vec{t})$ into
$\exists \vec{v}_{j1} \; \vec{v}_{j1} = \vec{t} \land R_j(\vec{v}_{j1})$ and
atoms $\overline{R_j}(\vec{t})$ into
$\exists \vec{w}_{j1} \; \vec{w}_{j1} = \vec{t} \land
   \overline{R_j}(\vec{w}_{j1})$.
Let
\[
\gamma_{j,m,m'} =
   \bigwedge_{k=1}^{m} R_j(\vec{v}_{jk}) \land
   \bigwedge_{k=1}^{m'} \overline{R_j}(\vec{w}_{jk}).
\]

A subformula $\psi_1 \lor \psi_2$ (resp., $\psi_1 \land \psi_2$) with
\[
\psi_i = \exists \vec{v} \vec{w} \;\;
   \alpha_i \land \bigwedge_{j=1}^s \gamma_{j,n_{ij},n'_{ij}}
\]
is turned into
\[
\exists \vec{v} \vec{w} \;\;
   \alpha \land \bigwedge_{j=1}^s \gamma_{j,m_j,m'_j}
\]
where
$m_j = \max(n_{1j}, n_{2j})$, $m_j' = \max(n_{1j}', n_{2j}')$ and
$\alpha = \alpha_1 \lor \alpha_2$
(resp.,
$m_j = n_{1j} + n_{2j}$, $m_j' = n_{1j}' + n_{2j}'$,
$\alpha = \alpha_1 \land \alpha_2'$, and $\alpha_2'$ is obtained from
$\alpha_2$ by replacing each variable $v_{jkl}$ by $v_{j(k+n_{1j})l}$ and
each variable $w_{jkl}$ by $w_{j(k+n_{1j})l}$).

For the equivalence of the rewritten formula to $\phi$, it is only necessary
to point out that since all the relations $R_j$ and $\overline{R_j}$ are
nonempty, $\psi_i$ is equivalent to
\[
\exists \vec{v} \vec{w} \;\;
   \alpha_i \land \bigwedge_{j=1}^s \gamma_{j,m_j,m'_j}.
\]

It is not hard to verify that the translation can indeed be implemented to
run in linear time and that the rewritten formula is of the form claimed in
the lemma.
\punto


\begin{theorem}
\label{theo:quant_free}
For any quantifier-free formula there is an equivalent expression
in WSA over universe relations and indicator relations
in which each indicator relation only occurs once.
\end{theorem}

\noindent {\bf Proof Sketch.}
Assume $R_1, \dots, R_s$ are all the predicates occurring in the formula.
By Lemma~\ref{lem:normal_form}, we only need to consider formulae of
syntax
\[
\phi = \exists \vec{v} \vec{w} \; \alpha \land
    \bigwedge_{j=1}^s \bigwedge_{k=1}^{m_j}  R_j(\vec{v}_{jk}) \land
                      \bigwedge_{k=1}^{m_j'} \neg R_j(\vec{w}_{jk})
\]
where $\alpha$ does not contain relational atoms
if each relation $R_j$ is nonempty and different from $U_j$.
Such a formula $\phi$ is equivalent to
\[
\exists \vec{v} \, \vec{w} \, \vec{t} \, \vec{t'} \; \alpha' \land
    \bigwedge_{j=1}^s \bigwedge_{k=1}^{m_j} 1_{R_j}(\vec{v}_{jk}, t_{jk}) \land
                      \bigwedge_{k=1}^{m_j'} 1_{R_j}(\vec{w}_{jk}, t_{jk}')
\]
with
\[
\alpha' = \alpha \land \bigwedge_{j=1}^s \Big(
                       \bigwedge_{k=1}^{m_j} t_{jk}  = 1
                 \land \bigwedge_{k=1}^{m_j'} t'_{jk} = 0 \Big).
\]
This is true because $R_j(\vec{v}_{jk})$ is equivalent to
$1_{R_j}(\vec{v}_{jk},1)$ and
$\neg R_j(\vec{w}_{jk})$ is equivalent to $1_{R_j}(\vec{w}_{jk},0)$.
Obtaining formulae of this form is indeed feasible because
$1_{R_j} \neq \emptyset$ and $1_{R_j} \neq U_j$.

Let $\vec{x}$ be the free variables of the formula. The WSA expression is
\[
\pi_{\vec{x}}(\sigma_{\alpha'}( B_1 \times \dots \times B_s ))
\]
with
\[
B_j :=
\rho_{\vec{v}_{j1}t_{j1}  \dots \vec{v}_{jm_j}t_{jm_j}
      \vec{w}_{j1}t'_{j1} \dots \vec{w}_{jm'_j}t'_{jm'_j}
} \big(
   (1_{R_j})^{m_j + m'_j}_{U_j}
\big).
\]
Each $B_j$ computes an $(m_j + m'_j)$-times product of $1_{R_j}$ using the
technique of Lemma~\ref{lem:kprod} which just uses one occurrence of $1_{R_j}$.
All the relations $1_{R_j}$ only occur once. This proves the theorem.
\punto

\begin{example}
\em
Consider an alternative encoding of 3-colorability in WSA which is based on
guessing a subset of relation $U = V \times \rho_C(\{r,g,b\})$.
Then 3-colorability is the problem of deciding the SO sentence
$
\exists C (\subseteq U)\; \neg \exists v, w, c, c' \;
   \phi_1 \lor \phi_2 \lor \phi_3
$
with
$\phi_1 = E(v,w) \land C(v,c) \land C(w,c)$,
$\phi_2 = C(v,c) \land C(v,c') \land c \neq c'$, and
$\phi_3 = \neg C(v,r) \land \neg C(v,g) \land \neg C(v,b)$,
i.e.,
$\phi_1$ asserts that two neighboring nodes have the same color,
$\phi_2$ that a node has simultaneously two colors, and
$\phi_3$ that a node has not been assigned any color at all.
If neither is the case, we have a 3-coloring of the graph.
Using Theorem~\ref{theo:quant_free}, $\phi_1 \lor \phi_2 \lor \phi_3$ becomes
\[
\pi =
   (\psi_1 \lor \psi_2 \lor \psi_3) \land
   1_C(u_1,c_1,t_1) \land 1_C(u_2,c_2,t_2) \land 1_C(u_3,c_3,t_3) \land
   1_E(v,w,t_4)
\]
where
\begin{eqnarray*}
\psi_1 &=& \; u_1=v \land u_2=w \land c_1=c_2 \land t_1=t_2=t_4=1
\\
\psi_2 &=& \; u_1=u_2 \land c_1 \neq c_2 \land t_1=t_2=1
\\
\psi_3 &=& \; u_1=u_2=u_3 \land c_1=r \land c_2=g \land c_3=b \land t_1=t_2=t_3=0;
\end{eqnarray*}
Following Theorem~\ref{theo:quant_free},
formula $\pi$ can be turned into WSA as
\[
Q_\pi := \sigma_{\psi_1 \lor \psi_2 \lor \psi_3}(
    \rho_{u_1 c_1 t_1 u_2 c_2 t_2 u_3 c_3 t_3}(
       (1_C)^3_{V \times \{r,g,b\}}
) \times \rho_{v w t_4}(E))
\]
where $(1_C)^3_{V \times \{r,g,b\}}$ denotes the WSA expression for
$1_C \times 1_C \times 1_C$ from Lemma~\ref{lem:kprod}.

The complete SO sentence can be stated as
\[
\exists 1_C \; (1_C: V \times \{r,g,b\} \rightarrow \{0,1\}) 
\land \neg \exists u_1 c_1 t_1 u_2 c_2 t_2 u_3 c_3 t_3 v w t_4 \; \pi.
\]
If
$1_C$ in $Q_\pi$ is replaced by
$\rk_{V,C}(V \times \rho_C(\{r,g,b\}) \times \rho_T(\{0,1\}))$,
this sentence can be turned into WSA without definitions as
$
\poss(\{\tuple{}\} - \pi_\emptyset(Q_\pi)).
$
\punto
\end{example}

\subsection{Quantification and Alternation}

Conceptually, in SO, there is no difference in the treatment of
second-order variables and relations coming from the input structure;
an existential second-order quantifier extends the structure over which
the formula is evaluated. In our algebra, however, we have to construct
the possible alternative relations for a second-order variable $R$ at the
beginning of the bottom-up evaluation of the algebra expression using
$\rk$ and have to later test the existential quantifier $\exists R$ using
the possible operation grouping the possible worlds that agree on $R$.
For that we have to keep $R$ around during the evaluation of the algebra
expression. Selections also must not actually remove tuples because this
would mean that the information about which world the tuple is missing from
would be lost. For example, the algebra expression corresponding to a
Boolean formula must not return false, but in some form must compute
the pair $\tuple{R, \mbox{false}}$.

Let $\phi$ be an SO formula with free second-order variables
$R_1, \dots, R_k$ and free first-order variables $x_1, \dots, x_l$.
Conceptually, our proofs will produce a WSA expression for $\phi$ that
computes, in each possible world identified by choices of relations
$R_1, \dots, R_k$ for the free second-order variables, the relation
\[
R_1 \times \dots \times R_k \times \Theta
\]
where $\Theta$ is a representation of a mapping
\[
\vec{a} \mapsto \mbox{truth value of } \phi[\vec{x}
   \mbox{ replaced by } \vec{a}].
\]
Truth and falsity cannot be just represented by 1 and 0,
respectively,
because an existential first-order quantifier will effect a projection
on $\Theta$ whose result may contain both truth values 1 and 0 for a
variable assignment $\vec{a}$.
Thus, projection may map environments for which $\phi$ is true together with
environments for which $\phi$ is false. In that case we would like
to remove the tuples for which the truth value encoding is 0.
Unfortunately, the function
\[
F:
\left\{ \begin{array}{l}
\{0\} \mapsto \{0\} \\
\{1\} \mapsto \{1\} \\
\{0,1\} \mapsto \{1\} \\
\end{array} \right.
\]
is nonmonotonic, and by Proposition~\ref{prop:mono} cannot be expressed in relational
algebra if the input relation is to occur in the query only once.
Fortunately, we do not need such a function $F$.

\begin{definition}
A PBIT (protected bit)
is either $\{\bot\}$ (denoting 0) or $\{\bot,1\}$ (denoting 1).
\end{definition}

Given a Boolean query $Q$ (i.e., $Q$ returns either $\{\tuple{}\}$ or
$\emptyset$),
\[
PBIT(Q) := (Q \times \{1\}) \cup \{\bot\}.
\]
The negation of PBIT $B$ is obtained by $\{\bot,1\} - (B \cap \{1\})$.
The set union on PBITs effects a logical OR, thus a relation
$\subseteq R \times PBIT$ for which $\tuple{\vec{a}, 1} \in R$ implies
$\tuple{\vec{a}, \bot} \in R$ guarantees that projecting away a column other
than the rightmost corresponds to existential quantification.

For an SO formula $\phi$ with free second-order variables
$R_1, \dots, R_k$ and free first-order variables $x_1, \dots, x_l$,
we will define a WSA expression that computes the relation
\[
TT(\phi) := 
1_{R_1} \times \dots \times 1_{R_k} \times \Theta
\]
such that
$
\Theta = (D^l \times \{\bot\}) \cup
\{ \tuple{\vec{a}, 1} \mid \phi[\vec{x} \mbox{ replaced by } \vec{a}] \mbox{ is true} \}
$
and
$D$ is a domain relation containing the possible values for the first-order
variables.
(So $\Theta$ can be thought of as a mapping $D^l \rightarrow PBIT$.)
The complement of such a relation $\Theta$ is
\[
\compl_{D^l}(\Theta) := D^l \times \{\bot,1\} - \sigma_{T=1}(\Theta).
\]

Next we obtain an auxiliary construction for
complementing a $\Theta$ relation while passing on the second-order relation.
This will be the essential tool for alternation.

\begin{lemma}
\label{lem:Theta_compl}
Let $P=1_{R_1} \times \dots \times 1_{R_k} \times \Theta$ where
$\Theta \subseteq D_1 \times \dots \times D_l \times PBIT$.
There is a WSA expression without definitions for
\[
\compl_{U_1, \dots, U_k;\vec{D},T}(S) :=
   1_{R_1} \times \dots \times 1_{R_k} \times \compl(\Theta)
\]
in which $P$ only occurs once.
\end{lemma}

\noindent {\bf Proof}.
Let $sch(U_i)=A_i$ and $sch(1_{R_i}) = A_i B_i$.
We write $\vec{1}$ for $1_{R_1} \times \dots \times 1_{R_k}$ and
$\vec{U}^+$ for
$U_1 \times \dots \times U_k \times \rho_{B_1 \dots B_k}(\{0,1\}^k)$.
A definition of $\compl_{U_1,\dots,U_k}(\vec{1})$ was given in
Section~\ref{subsect:ind}.
\begin{eqnarray*}
\compl_{U_1,\dots,U_k;\vec{D},T}(\vec{1} \times \Theta)
&=& \vec{1} \times (D^l \times \{\bot,1\} - \sigma_{T=1}(\Theta))
\\
&=& (\vec{U}^+ \times D^l \times \rho_T(\{\bot,1\}))
\\ 
&&  \hspace{5mm} - \compl_{U_1,\dots,U_k}(\vec{1}) \times D^l \times \rho_T(\{\bot,1\})
\\
&&  \hspace{5mm} - \vec{U}^+ \times \sigma_{T=1}(\Theta)
\\
&=& (\vec{U}^+ \times D^l \times \rho_T(\{\bot,1\})) 
\\
&& - \pi_{A_1,B_1,\dots,A_k,B_k,T} (
    \sigma_{\bigvee_i (A_i=A'_i \land B_i \neq B'_i) \lor T'=T=1} (
\\
&&  \hspace{5mm}
       \vec{U}^+ \times
       \rho_{A'_1 B'_1 \dots A'_k B'_k T'}(\underbrace{\vec{1} \times \Theta}_{P})
       \times \rho_T(\{\bot,1\}) ) ).
\end{eqnarray*}
The final WSA expression is in the desired form.
\punto

\medskip

Now we are ready to prove the main result of this section.

\begin{theorem}
Given a formula in second-order logic, an equivalent
WSA expression without definitions can be computed in linear time in
the size of the formula.
\end{theorem}

\noindent {\bf Proof Sketch.}
The proof is by induction.
Given second-order formula $\phi$ with free first-order variables $\vec{x}$
and zero or more free second-order variables.
%

Induction start: Assume that $\phi$ is quantifier-free.
Consider the quantifier-free formula
\[
\psi(\vec{x}, \vec{y}, t) :=
\Big( \bigwedge_{j:\; R_j \;\mathrm{is\; an\; SO\; var.}} R_{j}(\vec{y}_j)
\Big) \land \big( \phi \lor t=\bot \big),
\]
where the variables $\vec{y}$ and $t$ are new and do not occur in $\phi$.
It is easy to verify that $\psi$ defines the relation $TT(\phi)$.
Specifically, 
the projection down to columns $\vec{y}_j$ represents the
free second-order variable $R_j$, the projection down to columns
$\vec{x}$ specifies all the possible assignments to the first-order variables,
and $t$ is a PBIT for the truth value of $\phi$ for a given assignment
to the first- and second-order variables.
The corresponding WSA expression without definitions is
obtained using Theorem~\ref{theo:quant_free}.

Induction step ($\phi$ has quantifiers):
We assume that universal quantifiers $\forall \cdot$ have been replaced
by $\neg \exists \cdot \neg$.
Let $P$ be the WSA expression for $\psi$ claimed by the theorem.
\begin{itemize}
\item
First-order existential quantification:
If $\phi = \exists x_l \; \psi$, the corresponding WSA expression is
$\pi_{sch(P) - x_l}(P)$. It is easy to verify that the projection has exactly
the effect of existential first-order quantification,
$TT(\exists x_l \; \psi) = \pi_{sch(P) - x_l}(TT(\psi))$.

\item
Second-order existential quantification:
Let $R_1, \dots, R_s$ be the free second-order variables in $\psi$.
We may assume w.l.o.g.\ that these have disjoint schemas.
If $\phi = \exists R_s \; \psi$, the corresponding WSA expression is
\[
\pi_{sch(P) - sch(R_j)}(\poss_{sch(R_1), \dots, sch(R_{s-1})}(P)).
\]
Again, the correctness is straightforward,
$TT(\exists R_s \; \psi) = \pi_{sch(P) - sch(R_j)}(TT(\psi))$.

\item Negation:
By Lemma~\ref{lem:Theta_compl}, the WSA expression
$
\compl_{U_1 \dots U_s;\vec{D},T}(P)
$
is equivalent to $\phi = \neg \psi$.
\end{itemize}

All that is left to be done is to provide WSA expressions for the indicator
relations $1_{R_j}$.
For database relations $R_j$, the algebra expression is $\makeind(R_j, U_j)$.
For second-order variables $R_j$, it is $\rk_{sch(U_j)}(U \times \{0,1\})$.

For an SO sentence $\phi$ (i.e., without free variables), the algebra
expression computes a PBIT $TT(\phi)$ and its truth value is obtained
as $\pi_\emptyset(\sigma_{T=1}(\cdot))$.
\punto

\begin{example}
\label{ex:QBF2}
\em
We continue Example~\ref{ex:QBF1}.
Let
\[
\phi = \big( L(c,p,0) \land (P_1(p) \lor P_2(p)) \big) \lor
       \big( L(c,p,1) \land \neg (P_1(p) \lor P_2(p)) \big).
\]
Then $\Sigma_2$-QBF can be expressed by the SO sentence
\[
\exists P_1 (\subseteq V_1) \; \neg \exists P_2 (\subseteq V_2) \;
   \neg \exists c (\in C) \; \neg \exists p \; \phi.
\]

We can turn
\[
\big( \phi \lor t=\bot \big) \land P_1(p_{12}) \land P_2(p_{22})
\]
into WSA over indicator relations as
\[
Q = \sigma_\psi \big( \rho_{cpst_L}(1_L) \times
                \rho_{p_{11}t_{11}p_{12}t_{12}}((1_{P_1})^2_{V_1}) \times
                \rho_{p_{21}t_{21}p_{22}t_{22}}((1_{P_2})^2_{V_2}) \times
                \rho_{t}(\{\bot,1\}) \big)
\]
where
$
\psi \;=\; \big( t=\bot \lor (t_L = 1 \land p=p_{11}=p_{21} \land
           ((s=0 \land (t_{11}=1 \lor t_{21} = 1)) \lor
            (s=1 \land t_{11} \neq 1 \land t_{21} \neq 1)) ) \big).
$
Note that we have simplified the expression of the proof
somewhat by inlining the auxiliary variables $\vec{v}$ and $\vec{w}$.

The complete WSA expression for the SO sentence is
\begin{multline*}
\overbrace{\pi_\emptyset \circ \sigma_{t=1}}^{PBIT \;\mathrm{to\; bool}} \circ 
\overbrace{\pi_t \circ \poss}^{\exists P_1} \circ
\overbrace{\mbox{compl}_{V_1;T}}^{\neg} \circ
\overbrace{\pi_{p_{12} t_{12} t} \circ \poss_{p_{12} t_{12}}}^{\exists P_2}
\; \circ \\
\underbrace{\mbox{compl}_{V_1, V_2;T}}_{\neg} \; \circ
\underbrace{\pi_{p_{12} t_{12} p_{22} t_{22} t}}_{\exists c} \circ
\underbrace{\mbox{compl}_{V_1, V_2; C, T}}_{\neg} \circ
\underbrace{\pi_{p_{12} t_{12} p_{22} t_{22} c t}}_{\exists p}
(\underbrace{Q}_\phi).
\end{multline*}
We replace $1_L$ by $\makeind(L, \cdot)$ and $1_{P_i}$ by
$\rk_{p}(\rho_{pt}(V_i \times \{0,1\}))$.
\punto
\end{example}

Thus, definitions add no power to WSA.

\begin{corollary}
WSA without definitions captures WSA.
\end{corollary}

The data complexity of a query language refers to the problem of evaluating
queries on databases assuming the queries fixed and only the database part of
the input, while combined complexity assumes that both the query and the
database are part of the input \cite{Var82}.
Since SO logic is complete for the polynomial hierarchy (PHIER)
with respect to
data complexity and PSPACE-complete with respect to combined complexity
\cite{Sto1977}, a generalization of Fagin's Theorem \cite{Fag1974}
(see also \cite{Lib2004}),

\begin{corollary}
\begin{enumerate}
\item
WSA with or without definitions is
PHIER-complete with respect to data complexity,

\item
WSA with definitions is
PSPACE-hard with respect to combined complexity, and

\item
WSA without definitions is
PSPACE-complete with respect to combined complexity.
\end{enumerate}
\end{corollary}

We cannot directly conclude an
upper bound on the combined complexity of WSA with definitions
from the reduction of Theorem~\ref{theo:wsa2so} because it
was exponential-time: In the case that WSA definitions are used,
several copies of
formulae $\psi_V$ may be used in the SO formula constructed in the
proof, and that recursively.
However, we can think of the proof construction as a linear-time mapping
from WSA with definitions to second-order logic with definitions.
But the standard PSPACE algorithm for second-order logic extends directly
to second-order logic with definitions: Of the formula, we only have to
maintain a current path in its parse tree, which is clearly of
polynomial size. It follows that

\begin{proposition}
WSA with definitions is PSPACE-complete with respect to combined complexity.
\end{proposition}

\section{Related Work}
\label{sect:related}

In an early piece of related work, Libkin and Wong \cite{LW1996}
define a query algebra for handling both nested data types
and uncertainty. Their notion of uncertainty called {\em or-sets}\/
(as a generalization of the or-sets of \cite{INV1991}) is treated as
a special collection type that can syntactically be thought of as a
set of data and is only interpreted as uncertainty on an additional
``conceptual level''. The result is a very elegant and clean
algebra that nicely combines complex objects with uncertainty.
While their language is stronger and can manage nested data, there is
nevertheless a close connection to WSA, which can be thought of
as a flat relational
version of their language. Indeed, the or-set language contains
an operator $\alpha$ that is essentially
equivalent to the $\rk$ operator of WSA.

TriQL, the query language of the Trio project \cite{triobook2008},
subsumes the power of relational algebra and
supports an operation ``groupalts'' which expresses
the $\rk$ operation of WSA applied to a certain relation.
There are many more operations in TriQL, but it is hard to tell whether
$\poss_{\vec{A}}$ is expressible in TriQL since
no formal semantics of the language is available.
Moreover, TriQL contains a number of representation-dependent
(non-generic \cite{AHV95}) operations which
may return semantically different results for different semantically
equivalent representations of a probabilistic database. This makes
TriQL hard to study and compare with WSA. However, it seems that WSA is a good
candidate for a clean core to TriQL, and the results of the present paper
provide additional evidence that it is highly expressive.

The probabilistic databases definable using $\rk$ from certain relations
are also exactly the block independent-disjoint (BID) tables of
R\'e and Suciu. In their paper \cite{RS2007}, they study the related
representability problems for BID tables. Their results suggest that BID tables
are more powerful than tuple-independent tables, which correspond to
uncertain tables definable using the $\pws$ operation. This is in line
with observations made in Section~\ref{sect:weak} of the present paper.


The algebra defined in our own earlier work \cite{AKO07ISQL} is exactly the one
described in the present paper, modulo the following details. Most importantly,
while $\rk$ is introduced there as part of the algebra, most of the paper
focuses on the fragment that is obtained by replacing $\rk$ by choice-of.
Moreover, the syntax of $\poss_{\vec{A}}$ allows for the grouping of worlds
by a query $Q$ that can be given as a parameter; the syntax is $\poss_Q(Q')$.
An operation $\poss_{\vec{A}}$ in the syntax of the present paper corresponds
to an operation $\poss_{\pi_{\vec{A}}}$ in the syntax of \cite{AKO07ISQL}.
The results of this paper imply that allowing general queries $Q$ for grouping
adds no power, so we are indeed studying the same language.
The paper \cite{AKO07ISQL} also gives an SQL-like syntax for WSA, in
which the intuition of $\poss_{\vec{A}}$ is made explicit by the syntax
``select possible $\dots$ group worlds by $\dots$''.

In recent work \cite{AJKO2008,KO2008,Koch2008}, we have developed
efficient techniques for processing a large part of WSA. The only
operations that currently defy good solutions are $\poss_{\vec{A}}$
(i.e., with grouping, not $\poss_\emptyset$)
and, to a lesser extent, relational difference.
Indeed, the $\rk$ operator on the standard
representations described in Example~\ref{ex:stdrep} can be implemented
efficiently, even though semantically it generally causes an exponential blowup
in the size of the set of possible worlds.
Thus, it is natural to ask for the expressive
power of WSA with $\poss_{\vec{A}}$ replaced by $\poss$.
The construction of the proof of Theorem~\ref{theo:SO2WSAwdef} 
can map any SO formula of the form $\exists R \; \phi$ or $\forall R \; \phi$
where $\phi$ is FO to WSA. It is not hard to see that
despite the restriction to a {\em single}\/ second-order quantifier,
this fragment of WSA (with definitions) can express all of
NP $\cup$ co-NP.
For an upper bound, it seems that all such restricted WSA queries
have data complexity in $\Delta^P_2$ (i.e., $P^{\textit{NP}}$).

%

\section{Conclusions}
\label{sect:conclusion}

The main contribution of this paper is to give the apparently first
compositional algebra that exactly
captures second-order logic over finite structures, a logic of wide interest.

Second-order logic is a natural yardstick for the
expressiveness of query languages for uncertain databases. It is an
elegant and well-studied formalism that naturally captures what-if queries.
%
It can be argued that second-order logic takes
the same role in uncertain databases that
first-order logic and relational algebra
take in classical relational databases.  In that sense,
the expressiveness result of this paper, $WSA=SO$,
is an uncertain databases analog of Codd's Theorem.

Finding the right query algebra for uncertain databases is important
because efficient query processing techniques are easier to
obtain for algebraic languages without variables or quantifiers, and
algebraic operators are natural building blocks for database query plans.
Of course, the expressiveness result of this paper also implies that WSA
has high complexity and thus this paper can only be an initial call for the
search for more
efficiently processible fragments of WSA that retain some of its flavor of
simplicity and cleanliness.

\bibliographystyle{abbrv}
\bibliography{../bibtex}

\end{document}